\documentclass[12pt]{iopart}
\usepackage{amssymb}

\usepackage{epsfig}
\usepackage{dcolumn}
\usepackage{bm}
\usepackage{hyperref}
\usepackage{latexsym}

\begin{document}

\title{Cluster growth and dynamic scaling in a two-lane driven diffusive
system }
\author{I. T. Georgiev$^{1,2}$, B. Schmittmann$^{1}$ and R. K. P. Zia$^{1}$}
\address{$^{1}$Center for Stochastic Processes in Science and Engineering, 
Department of Physics, Virginia Tech, Blacksburg, VA 24061-0435, USA;\\
$^{2}$Integrated Finance Limited, 630 Fifth Avenue, Suite 450, New York, 
NY 10111, USA}
\date{January 18, 2006 }

\begin{abstract}
Using high precision Monte Carlo simulations and a mean-field theory, 
we explore coarsening phenomena in a simple driven diffusive system. The model
is reminiscent of vehicular traffic on a two-lane ring road. At sufficiently
high density, the system develops jams (clusters) which coarsen with time.
A key parameter is the passing probability, $\gamma$. For small values of 
$\gamma$, the growing clusters display dynamic scaling, with a 
growth exponent of 2/3. For larger values of $\gamma$, the growth
exponent must be adjusted, suggesting the ordered (jammed) state is not a 
genuine phase but rather a finite size effect.
\end{abstract}


\ead{\mailto{igeorgiev@iflt.com}, 
     \mailto{schmittm@vt.edu}, 
     \mailto{rkpzia@vt.edu}  }

\submitto{\JPA}

\section{INTRODUCTION}

Coarsening phenomena and pattern formation are fascinating subjects, 
drawing much attention especially in the context of first order
transitions \cite{Gunton,Langer,Bray}. Though these phenomena are
inherently dynamic in nature, there are intimate links to the statics
associated with thermal equilibrium. To be specific, the underlying rules of
evolution of this class of systems obey detailed balance (microscopic
reversibility); otherwise, these systems would never reach their equilibrium
states. More recently, considerable attention has been directed at a
different class of many-body systems, namely, those which settle into \emph{%
non-equilibrium }steady states. Unlike their equilibrium counterparts, these
steady states experience net \emph{fluxes} (of, say, energy or particles) 
\emph{through} the system. As a result, their stationary distributions lie
in general entirely outside the Boltzmann-Gibbs paradigm, and most are
completely unknown. Nevertheless, these systems are ubiquitous in nature 
(including, in particular, \emph{all} biological systems) and deserve 
our attention. The main distinguishing feature of non-equilibrium 
steady states is that they are coupled to more than one energy 
(or particle) reservoir, so that the underlying dynamics \emph{violates} 
detailed balance. Not surprisingly, such systems display much richer 
behavior than those in thermal equilibrium \cite{Beate_Royce_book, Mukamel_book}. 
Here, we ask a natural question: If such a system undergoes a first-order-like 
transition, will it also display very different coarsening phenomena? 
In other words, can we expect novel properties in systems evolving under 
a detailed balance violating dynamics?

Posed in this way, our question is extremely broad and progress towards an
answer will not be easy. One approach to this goal is to study idealized,
simple model systems. The hope is that, by exploring their behavior, we will
gain important insight towards the understanding of more complex real
systems. In addition, we may discover universality classes which encompass
both simple models and real systems. The Ising model played precisely such a
role in the understanding of thermodynamic singularities in an wide class of
equilibrium systems. In the same spirit, we explore coarsening phenomena in
a simple lattice gas model, driven into steady states far from thermal
equilibrium.

Our model is a member of a class known as driven diffusive systems.
Introduced by Katz, Lebowitz and Spohn \cite{KLS} as a seemingly trivial
modification of the Ising lattice gas, the prototype displayed many
surprising and counterintuitive phenomena. Since then, variations and
extensions of this simple model have been explored, and rich and complex
behavior continues to emerge \cite{Beate_Royce_book, Mukamel_book}. One
recent example is the drastic difference between a strictly one dimensional
system ($L\times 1$, also refered to as ``one-lane'') and a quasi-one
dimensional ($L\times 2$, ``two-lane'') lattice gas involving two species of
particles \cite{Korniss}. Driven in opposite directions on lattices with
periodic boundary conditions (``ring roads''), two particles encountering
one another may exchange positions with a small exchange rate $\gamma $. The
steady state of the \emph{one-lane} model is known analytically \cite{Sandow}
to remain homogeneous at all positive $\gamma $ and $\bar{\rho}$, the \emph{%
average }particle density or filling fraction. Meanwhile, Monte Carlo
studies of the \emph{two-lane} system showed a transition to an \emph{in}%
homogeneous state, with a single macroscopic cluster \cite{Korniss} (a
jammed state). Subsequently, simulations and a mean-field theory for a
very similar model also indicated such a kind of ``phase transition'' \cite%
{Arndt1,Arndt2}. 
Remarkably, an exact solution for this model was established later %
\cite{Rajewsky,Sasamoto}, showing that the jammed state is a finite-size
effect: Two-particle correlations are controlled by a \emph{finite} - but
gargantuan (e.g., $10^{70}$!) - length scale. Though there are
reasons to believe that our model also displays this type of finite-size
effect \cite{Kafri,Kafri1,Georgiev}, it is nevertheless interesting to
investigate jamming, and the growth of jams, in finite systems. In
particular, if the crossover to the asymptotic behavior does not occur until
the system sizes also reach $O\left( 10^{70}\right) $, then it is irrelevant
for all conceivable earthbound systems (e.g., pedestrian traffic and
biological motors).

With these considerations in mind, we explore coarsening phenomena in the
two-lane system, as it evolves toward a jammed state. In a recent study of
this model \cite{Mettetal,Beate}, clusters were found to grow much faster
than in typical non-driven diffusive systems \cite{Gunton,Langer,Bray}.
Here, we report an in-depth investigation, using both simulation and
analytic techniques. Focusing on dynamic scaling of the clustering process,
we investigate the average cluster size as a function of time and system
size. On the analytic side, we present a set of mean-field equations of
motion for the local particle densities. The numerical solutions of these
equations resemble the behavior of the exact solution of the deterministic
Burgers equation. From the Monte Carlo simulations, in the small $\gamma $
regime, we find a growth exponent of $2/3$ and good evidence of dynamical
scaling. Yet, as $\gamma $ increases, we are forced to adjust the scaling
exponents in order to collapse the data onto a universal curve. We believe
that these findings provide further evidence for the destabilization of the
ordered ``phase'' with increasing $\gamma$ or system size.

\section{The model and its properties}

The model is defined as a stochastic driven lattice gas with random
sequential dynamics that conserves the number of particles. There are two
kinds of particles, ``positive'' ($+$) and ``negative'' ($-$) ones, that
perform biased diffusion on an $L\times 2$ lattice with periodic boundary
conditions in both directions. Each site on the lattice can be empty ($%
\oslash $) or occupied by at most one (positive or negative) particle. There
are no other interactions between the particles except an excluded volume
interaction. The two species are driven in opposite directions by an
external field along the $x$-axis, where $x=1,...,L$ labels the position
of a site ``along the road''. The particles can also change lanes (with $%
y=1,2$ being the lane label). One Monte Carlo step (MCS) contains the
following steps repeated $2L$ times:

\begin{itemize}
\item randomly pick a bond connecting two nearest-neighbor sites on the
lattice;

\item if the bond is in the $x$-direction and connects either a $(+\oslash )$
or a $(\oslash -)$ pair, exchange the particle and the vacancy (a
particle-hole exchange). If it is a particle pair $(+-)$, then perform
the exchange with probability $\gamma $ - a process also referred to as
charge-exchange, or passing in the language of traffic models;

\item if the bond is in the transverse direction (cross-lane), then
exchange particle-hole pairs always and execute charge exchanges with
probability $\gamma $.
\end{itemize}

\begin{figure}[tbp]
\begin{center}
\epsfig{file=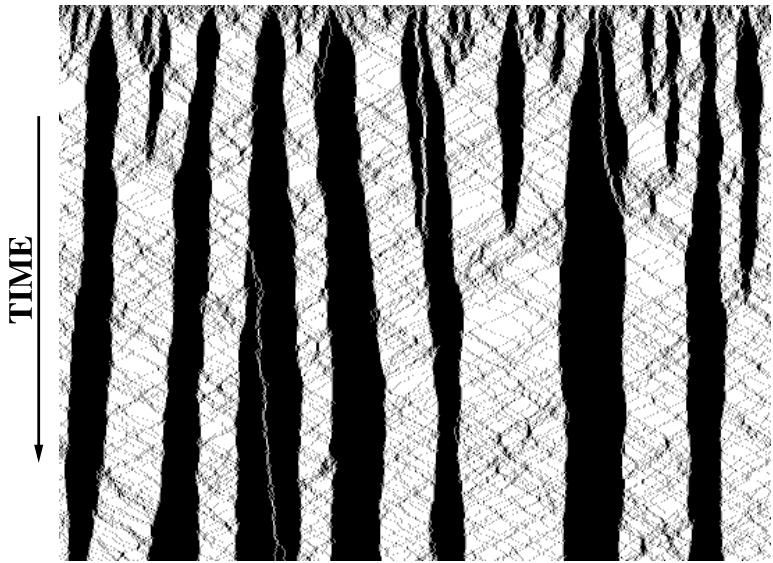,width=4.0in}
\caption{Space-time plot of a $512 \times 2$ lattice 
with particle density 
$\bar{\rho} =0.5$ and charge-exchange probability $\gamma =0.2$. 
The initial configuration is random. For each $x$, a black dot represents
particles in both lanes, a white dot represents no particles, and a gray dot
represents one of the two lanes being occupied.} 
\label{fig:L512_sptime}
\end{center}
\vspace{-0.7cm}
\end{figure}

\noindent Since no backward jumps are allowed, these dynamical rules can
be interpreted as imposing an infinite external ``electric field'' on the
``charged'' particles, which perform random walks \emph{biased} only along $%
x $. Apart from the random choosing of bonds, the only stochastic parameter
in the model is the particle-exchange probability $\gamma $. Our dynamics is
invariant under spatial translations and charge-parity (CP) transformations.
We have studied only the case of a neutral system, i.e. $N_{+}=N_{-}$, where 
$N_{+}$ and $N_{-}$ are the number of $+$ and $-$ particles respectively.
Further, we focus mostly on systems at half filling, i.e., $0.5=\bar{\rho}%
\equiv \left( N_{+}+N_{-}\right) /\left( 2L\right) $.

Unlike its equilibrium analogue, i.e., the Ising model, there is a
qualitative difference between the $L\times 2$ (quasi 1-D) and the $%
L\times 1$ (strictly 1-D) systems. For the latter case, both the exact
solution (for an infinite system) \cite{Sandow} and computer simulations (of
finite systems) \cite{MCref} show that the steady state is disordered and
homogeneous. Specifically, if we define a cluster as a collection of
particles connected by nearest-neighbor bonds, then the distribution of
cluster sizes is controlled by an exponential with a characteristic length
of $O\left( L^{0}\right) $, remaining finite except in the singular
limits $\gamma \rightarrow 0$ or $\bar{\rho}\rightarrow 1$. By contrast, the
presence of a second lane has a profound effect on the steady state. For
large regions in the $\gamma $-$\bar{\rho}$ plane, the cluster-size
distribution has another peak located at sizes $\propto L$, which is a clear
signal of a macroscopic jam \cite{Korniss,Mettetal,Beate,Georgiev}. This
behavior is very counterintuitive: Adding a second lane should not lead to
worse traffic jams! According to one's na{\"{\i}}ve expectation, the
extra degree of freedom associated with the additional lane should provide
an efficient mechanism for relieving blockages and allowing the system to
reach a more homogeneous state. Instead, computer simulations reveal
properties more akin to the 2-D system ($L\times L$), where a compact,
macroscopic cluster is present in significant portions of the $\gamma $-$%
\bar{\rho}$ phase diagram. Although the macroscopic cluster in the $L\times
2 $ case is unlikely to persist in the $L\rightarrow \infty $ limit \cite%
{Georgiev}, our focus here is how the system evolves toward a
phase-separated state in (physically reasonable) $L$'s up to 
$O(10^{5}).$

\begin{figure}[tbp]
\vspace{-0.3cm}
\begin{center}
\epsfig{file=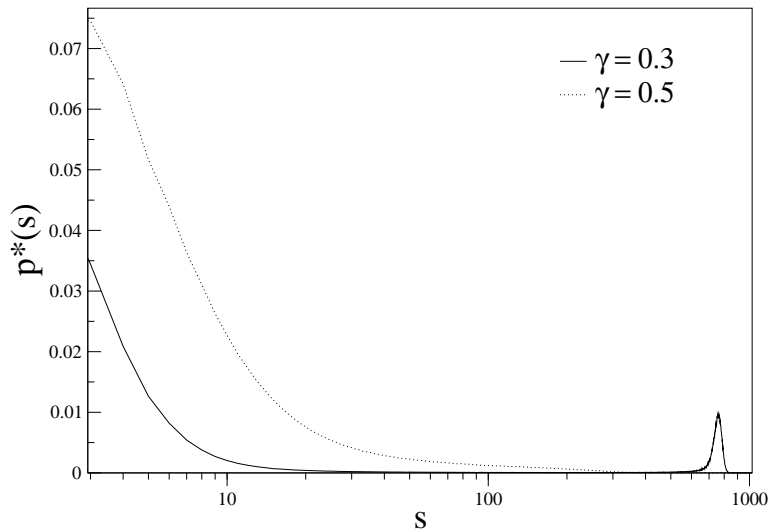,angle=-90, width=4.5in}
\vspace{-0.3cm}
\caption{The residence distribution for two values of $\protect\gamma $ on a 
$1024\times 2$ lattice, for $\bar{\rho}=0.5$. 
For $\protect\gamma =0.3$ we observe the ordered
phase with a bi-modal distribution (a decay for small values of $s$ and a
peak corresponding to the macroscopic cluster), while for $\protect\gamma %
=0.5$ we have a disordered phase with uni-modal residence distribution.  }
\label{res_dist}
\end{center}
\vspace{-0.5cm}
\end{figure}

Just like in the daily commute, for low $\gamma $ and moderate $\bar{\rho}$,
a local blockage will remain long enough for the next group of particles to
jam into it. As a result, the blockage grows, as long as the influx, due
to the arrival of particles from \emph{outside} the cluster, exceeds the
outflux, due to the loss of particles from non-zero $\gamma $. 
Fig.~\ref{fig:L512_sptime} provides an example of this growth 
process. It is a space-time
plot of a $512\times 2$ lattice starting from a random initial
configuration. The overall particle density is $\bar{\rho}=0.5$ (i.e. $\bar{%
\rho}_{+}=\bar{\rho}_{-}=0.25$) and the charge-exchange probability is $%
\gamma =0.2$. Each row on the figure encodes the microscopic configuration
on the lattice. Time runs vertically downwards, covering the first $400$
MCS. For these values of $L,\bar{\rho},$ and $\gamma $, the system will
coarsen until there is only one macroscopic cluster. The remainder of this
paper will be devoted to a more quantitative characterization of this
coarsening process. Before continuing, let us provide some necessary
definitions.

First, we define a cluster of size $s$ as any $s$ particles connected by
nearest-neighbor bonds regardless of their charge (i.e., a ``mass''
cluster). By collecting a histogram of cluster sizes formed at
time $t$, we can estimate the probability $\tilde{p}(s,t)$ of having a
cluster with a mass $s$ at time $t$. Since $\tilde{p}$ is not a conserved
distribution, it is customary to study the residence distribution, defined
as $p(s,t)\equiv s\;\tilde{p}(s,t)$. Representing the probability that a
randomly chosen particle belongs to a cluster of size $s$ at time $t$, this
distribution is easily normalized, i.e., $\sum_{s}p(s,t)=1$. The
steady-state residence distribution $p^{\ast }(s)$ is given by $p^{\ast
}(s)=\lim_{t\rightarrow \infty }p(s,t)$. In the 1-lane system, it is
monotonically decreasing and, in the thermodynamic limit, known analytically
: $p^{\ast }(s)\propto s^{-1/2}\exp (-s/\xi )$ \cite{Sandow}, where $\xi (%
\bar{\rho},\gamma )$ is a characteristic mass independent of $L$. However,
for the 2-lane model, $p^{\ast }$ changes from being monotonic for larger
values of $\gamma $ to being bi-modal for small $\gamma $ (and moderate $%
\bar{\rho},L$). An example is shown in Fig.~\ref{res_dist}, where $\bar{\rho}%
=0.5,\,L=1024,$ and $\gamma $ takes two values, $0.5$ and $0.3$. For the
larger $\gamma $, $p^{\ast }$ is similar to the 1-lane case. For the smaller
value, $p^{\ast }$ has a second peak at $s=O\left( L\right) $, associated
with the macroscopic cluster. Meanwhile, the peak at $s=0$ can be traced to
the region\emph{\ outside }this cluster, where a very low particle density
allows only for very small blockages ($s\leq 5$). We refer to the particles
in this region as ``travellers'' (see also Fig.~\ref{fig:L512_sptime}).
Further details about the stationary states and the $\gamma $-$\bar{\rho}$
phase diagram can be found in \cite{Georgiev}.

\section{Monte Carlo simulations}
\label{MC}

In a preliminary study of coarsening in this model \cite{Mettetal,Beate},
systems up to $L=10^{4}$ with $\gamma =0.1$ were simulated. The main
observation is that the cluster sizes appear to grow roughly as $t^{2/3}$.
Here, we report the results of a refined study using a fast multi-spin
coding algorithm where the storage of the microstates on the lattice and the
updating rules are coded in a bit-wise manner, as in, say, \cite{newman}.
Since our model is a three-state model, we need at least two bits to store
the three possible states on a single lattice site. Instead of using two
consecutive bits in a machine word for encoding the state on a site, we
prefer to use one bit from two different words. In this way the algorithm
for a single Monte Carlo step becomes somewhat simpler, though working with
two different arrays representing the two lanes reduces the efficiency
slightly. Each element of the two arrays represents a particular site on one
of the lanes. Exploiting the $64$-bit architecture of our hardware, the
individual bits in the machine word encode the state on this site for $64$
different lattices, all of which are updated simultaneously. The increase in
the efficiency of the multi-spin algorithm is roughly $35$-fold over the
standard single spin flip algorithm. This allows us to study much larger
systems and provides much better statistics.

\begin{figure}[tbp]
\begin{center}
\epsfig{file=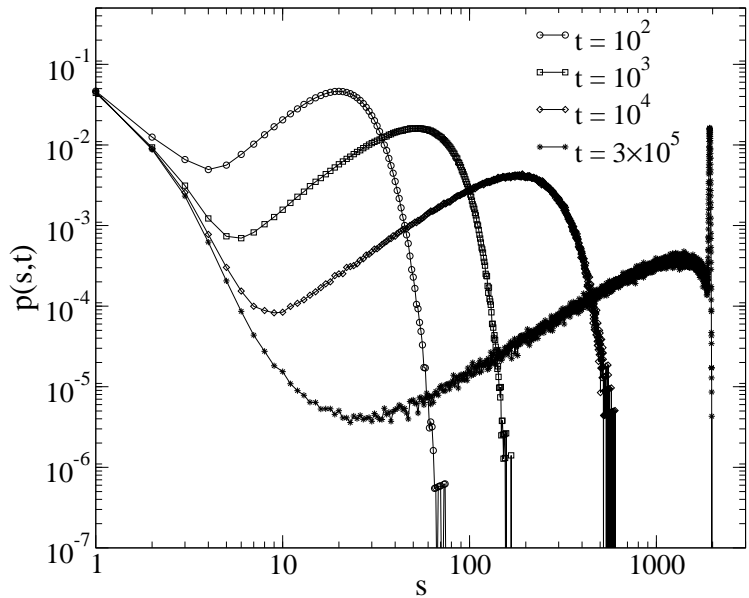,angle=-90, width=4.3in}
\caption{The residence distribution $p(s,t)$ \textit{vs} the cluster 
mass $s$ at
different times $t=10^2, 10^3, 10^4$ and $3\times10^5$ MCS. The lattice size
is $L=2048$ with $\gamma=0.1$ and $\bar{\rho}=0.5.$ }
\label{res_distr_time}
\end{center}
\vspace{-0.5cm}
\end{figure}

Fig.~\ref{res_distr_time} shows the behavior of the residence distribution $%
p(s,t)$ on a $2048\times 2$ lattice with $\gamma =0.1$ and particle density $%
\bar{\rho}=0.5$. In the regime from $\sim 10^{2}$ to $\sim 10^{5}$ MCS, 
seemingly unbounded coarsening occurs. Thereafter, the largest
cluster saturates the system size and the system settles into a steady
state. Two components of $p(s,t)$ are easily distinguished, corresponding to
a relatively static mode near the origin associated with the travelers, 
and a more ``dynamic'' mode associated
with the growing clusters. Since these modes exhibit drastically different
time dependences, the full distribution cannot be collapsed on a single
universal curve. We can, however, isolate the ``growing'' component by
introducing a coarse-grained description of the growth process 
\cite{Mettetal}.

\begin{figure}[tbp]
\begin{center}
\epsfig{file=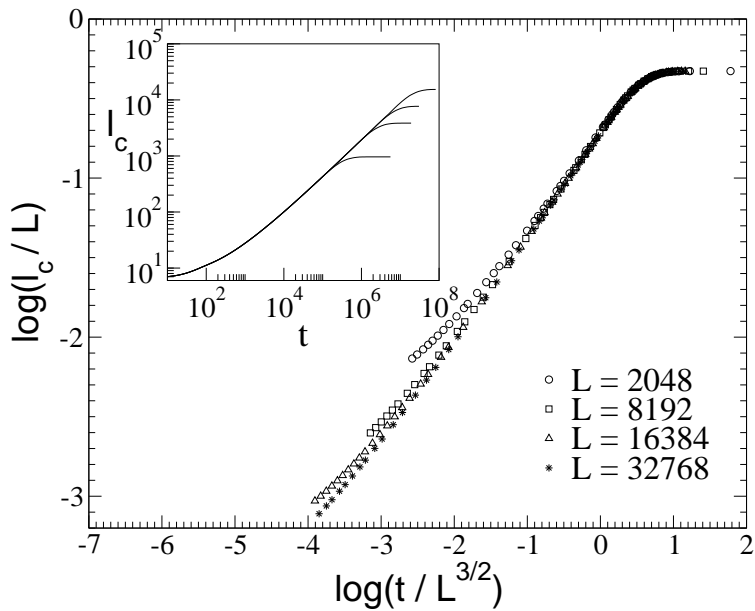,angle=-90, width=4.3in}
\caption{Dynamic scaling for $l_c/L^\protect\protect\alpha$ vs. $t/L^z$ for $%
\protect\gamma=0.1$ and $\bar{\rho}=0.5$. The graphs collapse on an
universal curve for large $t$ with $\protect\alpha=1.0$ and $z=3/2$. The
inset shows the original data $l_c$ vs. $t$, where the lattice sizes are $%
L=2048, 8192, 16384$ and $32768$ from top to bottom. }
\label{G01_growth_scaling}
\end{center}
\vspace{-0.5cm}
\end{figure}

\begin{figure}[tbp]
\begin{center}
\epsfig{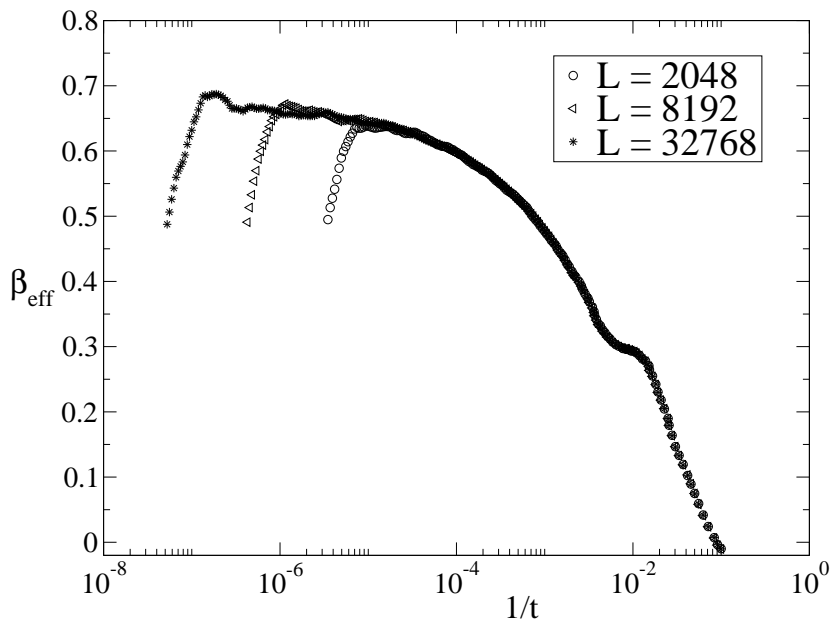}
\caption{The effective growth exponent $\beta_{eff}$ for 
$\gamma=0.1$ and lattice sizes $L=2048, 8192$ and $32768$. }
\label{G01_growth_exp}
\end{center}
\vspace{-0.5cm}
\end{figure}

\begin{figure}[tbp]
\begin{center}
\epsfig{file=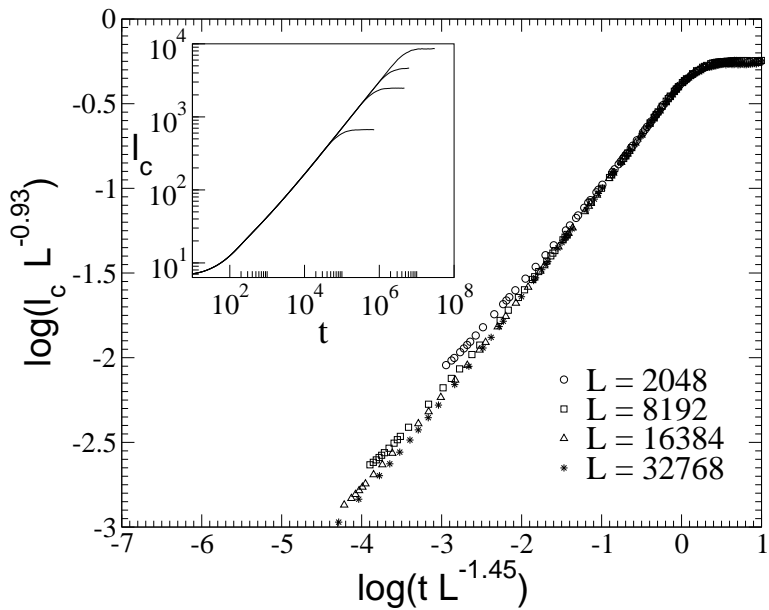,angle=-90, width=4.3in}
\caption{Dynamic scaling for $l_c/L^\protect\protect\alpha$ vs. $t/L^z$ for $%
\protect\gamma=0.3$ and $\bar{\rho}=0.5$. The graphs collapse on an
universal curve for large $t$ with $\protect\alpha=0.93$ and $z=1.45$. The
inset shows the original data $l_c$ vs. $t$, where the lattice sizes are $%
L=2048, 8192, 16384$ and $32768$ from top to bottom. }
\label{G03_growth_scaling}
\end{center}
\vspace{-0.5cm}
\end{figure}

\begin{figure}[tbp]
\begin{center}
\epsfig{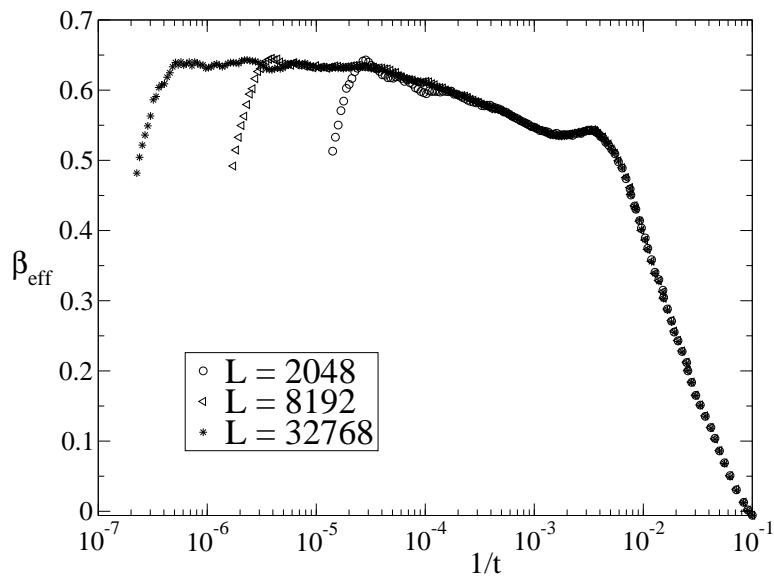}
\caption{The effective growth exponent $\beta_{eff}$ for $\gamma=0.1$ 
and lattice sizes $L=2048, 8192$ and $32768$. }
\label{G03_growth_exp}
\end{center}
\vspace{-0.5cm}
\end{figure}

We define a strictly one-dimensional coarse-grained configuration on a $%
L\times 1$ chain as follows:\ (i) for a suitably chosen positive integer $b$%
, center a $b\times 2$ window at each site $x=1,...,L$ with, say, $y=1$
of the original lattice and count the number of particles inside the window;
(ii) if this particle number exceeds $b$, we occupy the associated site $x$
on the (coarse-grained)\ $L\times 1$ lattice; otherwise, that site remains
empty. Clearly, this procedure is independent of $y$, i.e., which lane is
selected. Since the small clusters in the traveler region rarely exceed a
mass of $5$, we use windows of length $b=5$. As a result, the traveler
component is removed from the coarse-grained configuration which contains
only the information about the growth mode. Since large clusters are almost
unaffected by the coarse-graining procedure, it should be obvious that the
large-scale dynamics on the new $L\times 1$ lattice has the same
characteristics as the original system. The coarse-graining procedure is
performed each time we take data, and the information about the lengths of
clusters on the new lattice is collected. The coarse-grained residence
distribution, $p(l,t)$, on the new lattice is defined as before, namely, as
the (normalized) probability that a randomly chosen particle belongs to 
a cluster of length $l$ at time $t$. 
The average cluster length $l_{c}(t)$ is defined as
its first moment:\ 
\begin{equation}
l_{c}(t)\equiv \sum_{l=1}^{L}l\;p(l,t)
\label{l_c}
\end{equation}
In the following, we show two large data sets for $l_{c}$ vs. $t$ obtained
from $4096$ independent runs, on lattices with $L=2K,$ $8K,$ $16K$, and $32K$%
, and $\bar{\rho}=0.5$. One data set shows a relatively small value of
$\gamma $, $\gamma =0.1$ (Figs.~\ref{G01_growth_scaling} and 
\ref{G01_growth_exp}); the other shows the significantly larger
value $\gamma =0.3$ (Figs.~\ref{G03_growth_scaling} and \ref{G03_growth_exp}).
All initial ($t=0$) configurations are random.

To analyze the data, we test whether they satisfy dynamic scaling of the
form: 
\begin{equation}
l_{c}(t)=L^{\alpha }\;F(t/L^{z})\;  \label{dyn_scale}
\end{equation}%
This scaling form is chosen to reflect two features of our data. On the one
hand, on any given lattice, we expect a saturation of cluster sizes for
large times, so that $\lim_{t\rightarrow \infty }l_{c}(t)$ should become
independent of time. In other words, the scaling function $F(x)$ should
approach a constant as $x\rightarrow \infty $. Any $L$-dependence of the
steady-state cluster size is reflected in the exponent $\alpha $, via $%
\lim_{t\rightarrow \infty }l_{c}(t)\propto L^{\alpha }$. On the other hand,
we check whether the growth regime follows a power law, corresponding to $%
l_{c}(t)\propto t^{\alpha /z}$ and a finite limit of $x^{-\alpha /z}F(x)$
for $x\ll 1$. For $\gamma =0.1$, the data collapse is very good at large $t$%
, using the exponent values $\alpha =1.00 \pm 0.01$ and $z=1.50 \pm 0.01$.
This gives us a cluster growth exponent $\beta \equiv \alpha /z=0.66 \pm
0.02 $ consistent with a value of $2/3$. It is not surprising, of course,
that the scaling form of Eq.~(\ref{dyn_scale}) does not hold for early
times. Starting from random configurations, particles immediately follow
their preferred direction, rapidly generating many small clusters separated
by completely empty regions. The associated $l_{c}(t)$ is nearly constant
for some time and independent of system size, so that the corrections to
Eq.\ (\ref{dyn_scale}) decrease with $L$. The average cluster size only
starts to increase noticeably when particles begin to escape from these
small clusters, initiating the growth regime.
 
To put the spot light squarely on the growth regime, we show the local
exponent, $\beta _{eff}\equiv d\ln (l_{c}(t))/d(\ln t)$ in Fig.\
\ref{G01_growth_exp}, for a range of system sizes. This representation is of
course much more sensitive to small changes in the power law, $l_{c}(t)\propto
t^{\beta _{eff}}$, than Fig.\ \ref{G01_growth_scaling}. The data show that $%
\beta _{eff}$ is monotonically increasing with $t$ and independent of $L$
until the saturation regime sets in. The onset of saturation is marked by a
sharp spike in $\beta _{eff}$, followed by a similarly sharp drop. Just
before saturation, $\beta _{eff}$ approaches the limiting value $2/3$. The
apparent $t$-dependence of $\beta _{eff}$ simply reflects the behavior of
the scaling function $F(t/L^{z})$. The behavior of our system after
saturation, reflected in the exponent value $\alpha =1.0$ shows that the
largest clusters are indeed macroscopic, in that their size scales linearly
with $L$.

\begin{figure}[tbp]
\begin{center}
\epsfig{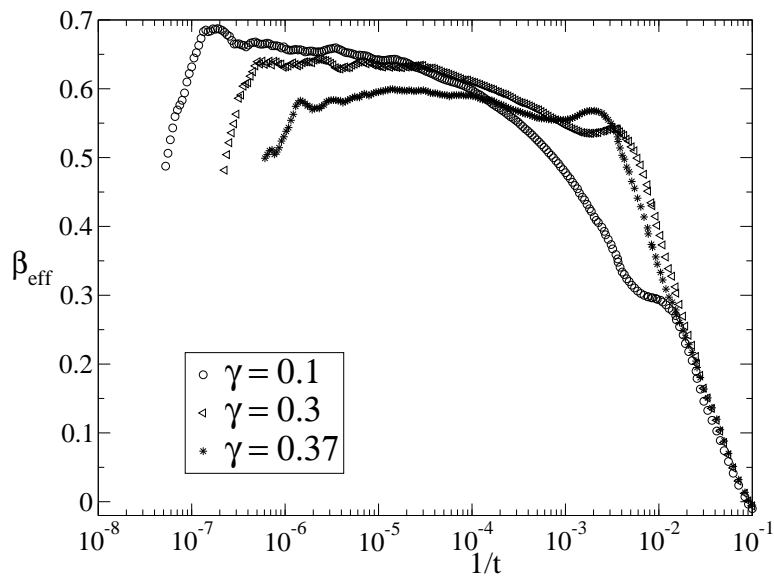}
\caption{The effective growth exponent $\beta_{eff}$ 
for $\gamma=0.1, 0.3$ and $0.37$ for
lattice size $L=32768$.}
\label{Growth_exp_L32768}
\end{center}
\vspace{-0.5cm}
\end{figure}

The results for the case of $\gamma =0.3$ are shown in 
Figs.~\ref{G03_growth_scaling}
and \ref{G03_growth_exp}. By increasing $\gamma $, we 
effectively destabilize the
macroscopic clusters in the stationary regime.
Here, the best data collapse is obtained with $\alpha =0.93\pm 0.01$ and $%
z=1.45\pm 0.01$, which results in a growth exponent $\beta =0.64\pm 0.02$.
The smaller value of $\beta $ is also clearly reflected in the plot of the
local slope, Fig.\ \ref{G03_growth_exp}. The need to adjust the scaling
exponents as functions of $\gamma $ may well point towards significant
changes in the steady-state properties of the model. 
For sufficiently large $L$ and
$\gamma $, a crossover to an exponential (rather than bi-modal)\
residence distribution occurs \cite{Georgiev}. For our 
parameters ($\bar{\rho}=0.5$,
$\gamma =0.3$, $L\leq 32k$), the stationary distribution is still
bi-modal but the peak at large $s$ is already very broad. While the data
collapse for the coarsening system is still very good, a value of $\alpha <1$
shows that the largest clusters already scale sublinearly with $L$. It is
conceivable that Eq.\ (\ref{dyn_scale}) acquires additional corrections
which account for the crossover, leading to effectively $\gamma $-dependent
exponents. For $\gamma =0.37$ (Fig.~\ref{Growth_exp_L32768}) 
it becomes blatantly
obvious that $\beta _{eff}$ remains well below $0.6$ and that there are
strong $L$-dependent corrections.

To summarize, for small values of $\gamma $ (i.e. $\gamma \leq 0.1$), the
system coarsens with a growth exponent $\beta $ consistent with $2/3$, and
approaches a steady state characterized by a single macroscopic cluster,
with $l_{c}\propto L$. As $\gamma $ increases, the data collapse remains
good but only if the exponents are adjusted with $\gamma $. In particular,
the growth exponent $\beta $ must be reduced quite significantly. It is
natural to conclude that this is a by-product of the crossover from bi-modal
to exponential $p^{\ast }(s)$.

\section{A mean-field approximation}

Noise plays a significant role in our Monte Carlo simulations. First, all
initial conditions are random; second, and more importantly, the dynamics
itself is inherently stochastic. Even starting from the same initial
condition, each Monte Carlo run follows a distinct trajectory. It is natural
to ask whether the statistical averaging over these different trajectories
is essential for the coarsening process. To put this question into context,
we recall that thermal fluctuations play no role for the temporal evolution
of equilibrium systems towards their final ordered states, provided the
domain size is much larger than the thermal correlation length \cite{Bray}.
As a result, domain growth laws can be obtained by integrating a
deterministic mean-field equation, e.g., the Cahn-Hilliard equation for the
ordering dynamics of a scalar conserved order parameter (Model B, in the
Hohenberg-Halperin classification \cite{HH}). 

In this section, we neglect fluctuations, in an attempt to test their
relevance. We establish evolution equations for the two particle densities
which form the analogue of the Cahn-Hilliard equation for our dynamics and
integrate them numerically. The only source of noise will be in the initial
condition; once it is set, the evolution proceeds completely
deterministically. If we recover a growth exponent of $2/3$, we may conclude
that the stochasticity of the Monte Carlo simulations is irrelevant. In
contrast, if a different growth exponent emerges here, we are faced with a
stark difference between coarsening towards an equilibrium vs a
non-equilibrium steady state, namely, that fluctuations are essential to
describe coarsening phenomena in the latter.

To obtain a set of evolution equations, we begin from the microscopic
master equation and derive equations for the local averages, $\left\langle
\rho ^{+}(x,y,t)\right\rangle $ and $\left\langle \rho
^{-}(x,y,t)\right\rangle $, of positive and negative particles at lattice
site $x,y$ and time $t$. Alternatively, we can write down simple balance
equations for the loss and/or gain of particles at a given lattice site.
Since the numerical integration will require a discrete spatial variable, we
simply work with the original lattice, rather than taking a naive continuum
limit. We assume spatial homogeneity in the $y$-direction, letting the
densities depend on $x$ only, $\left\langle \rho ^{+}(x,y,t)\right\rangle
\rightarrow \left\langle \rho ^{+}(x,t)\right\rangle $. In a mean-field
approximation, all higher-point averages are replaced by products of
one-point averages, e.g., $\left\langle \rho ^{+}(x+1,t)\rho
^{-}(x,t)\right\rangle \simeq \left\langle \rho ^{+}(x+1,t)\right\rangle
\,\left\langle \rho ^{-}(x,t)\right\rangle $. Finally, for simplicity, we
omit the $\left\langle \cdot \right\rangle $ brackets, and introduce the
local hole density: 
\begin{eqnarray}
\phi (x,t)\equiv 1-\rho ^{+}(x,t)-\rho ^{-}(x,t)
\nonumber
\end{eqnarray}%
The resulting equations, in one spatial dimension, are given by \cite%
{Korniss2}:\ 
\begin{eqnarray}
\partial _{t}\rho ^{+}(x,t) &=&\rho ^{+}(x-1,t)\phi (x,t)-\rho ^{+}(x,t)\phi
(x+1,t) \nonumber \\
&+&\gamma \left[ \rho ^{+}(x-1,t)\rho ^{-}(x,t)-\rho ^{+}(x,t)\rho
^{-}(x+1,t)\right]   \nonumber \\
\partial _{t}\rho ^{-}(x,t) &=&\rho ^{-}(x+1,t)\phi (x,t)-\rho ^{-}(x,t)\phi
(x-1,t) \nonumber \\
&+&\gamma \left[ \rho ^{-}(x+1,t)\rho ^{+}(x,t)-\rho ^{-}(x,t)\rho
^{+}(x-1,t)\right]   \label{MFT}
\end{eqnarray}%
Each of the terms on the right hand side is easily understood; for example,
the contribution $\rho ^{+}(x-1,t)\phi (x,t)$ reflects the jumps of positive
particles at site $x-1$ to site $x$, provided $x$ is empty. The equations
for $\rho ^{+}(x,t)$ and $\rho ^{-}(x,t)$ are of course related by a
charge-parity (CP) symmetry. The two equations take the form of continuity
equations, i.e.,
\begin{eqnarray}
\partial _{t}\rho ^{\pm}(x,t) &=&j^{\pm}(x-1,t)-j^{\pm}(x,t)  
\label{MFT-currents}
\end{eqnarray}%
where $j^{+}(x,t)\equiv \rho ^{+}(x,t)\left[ \phi (x+1,t)+\gamma \rho
^{-}(x+1,t)\right] $ and $j^{-}(x,t)\equiv -\rho ^{-}(x+1,t)\left[ \phi
(x,t)+\gamma \rho ^{+}(x,t)\right] $ denote the net currents of positive and
negative particles from site $x$ to site $x+1$, respectively.

Before we turn to the numerical integration of these equations, we first
summarize the key properties of their \emph{stationary} solutions, $\rho
^{+}(x,t)\rightarrow \rho ^{+}(x)$ and $\rho ^{-}(x,t)\rightarrow \rho
^{-}(x)$. This will help us interpret the density profiles at late times.
Since Eqns.~(\ref{MFT}) take the form of continuity equations, they are
clearly satisfied by spatially uniform densities, reflecting homogeneous
steady states. In the \emph{continuum} limit,
spatially inhomogeneous stationary solutions, reflecting jammed states,
can also be found \cite{Korniss2}. 
These consist of a domain of travellers (characterized by
uniform densities) and a single large cluster, containing essentially no
holes (i.e., $\phi \simeq 0$ inside), linked by a narrow region in which the
densities vary rapidly, reminiscent of shocks. Remarkably, it is
straightforward to find \emph{approximate} solutions to Eqns.~(\ref{MFT})
which consist of separate uniform and cluster domains, patched together by a
consistency condition. Stationarity implies $j^{\pm }(x,t)\equiv j^{\pm
}=const$ independent of $t$ and $x$. The CP symmetry also leads to $%
j^{+}=-j^{-}\equiv j$. In the traveller domain, all densities are uniform
whence $j=\rho ^{+}\left[ \phi +\gamma \rho ^{-}\right] $. Invoking symmetry
again, we expect $\rho ^{+}=\rho ^{-}\equiv \rho $, and $\phi =1-2\rho $ so
that we can find the current-density relation in the traveller domain:\ 
\begin{equation}
j=\rho \left[ 1-2\rho +\gamma \rho \right]   \label{MFT-trav}
\end{equation}%
Next, we center the $x$-axis in the middle of the cluster and simplify Eqns (%
\ref{MFT}) by setting $\phi (x)=0$, i.e., $1=\rho ^{+}(x)+\rho ^{-}(x)$. By
continuity, the currents inside the cluster must equal the exterior currents
whence \ 
\begin{equation}
j=\gamma \rho ^{+}(x)\rho ^{-}(x+1)=\gamma \rho ^{+}(x)\left[ 1-\rho
^{+}(x+1)\right]   \label{MFT-cluster}
\end{equation}%
This is a recursion relation for $\rho ^{+}(x)$, solved by
\begin{equation}
\rho ^{+}(x)=\frac{1}{2}\left[ 1-A\tan \mu x\right]   \label{rho+}
\end{equation}%
with the amplitude and slope of the $\tan $-function given in terms of the
excess current, $j_{ex}\equiv j-\gamma /4>0$, via \ 
\begin{eqnarray}
A=\tan \mu =\sqrt{j_{ex}}
\nonumber
\end{eqnarray}%
Clearly, the corresponding negative charge density is given by $\rho
^{-}(x)=1-\rho ^{+}(x)$. To match the solutions inside and outside the
clusters, we note that the positive particles form a sharp shock front at
the \emph{left} edge of the cluster while behaving quite smoothly at the 
\emph{right} edge \cite{Arndt3}. Assuming
the cluster ranges from $x=-l/2$ to $x=+l/2$, we match the interior density
to the exterior one by equating them at the \emph{right} edge:
\begin{eqnarray}
\rho =\rho ^{+}(l/2)
\nonumber
\end{eqnarray}%
To close the set of equations, one must add the overall mass constraint, 
$2\bar{\rho}L=l+2\rho (L-l)$ to these relations.

The two key results from this section are (i) the $\tan $-shape of the
density profiles inside a stationary cluster, and (ii) the excess current $%
j_{ex}$ through a cluster of length $l$. Instead of the lengthy full
expression, we only give its form for large $l$, namely, $j_{ex}\simeq \pi
^{2}\gamma /l^{2}$.

To solve the mean-field equations numerically, we use a simple Euler
integration scheme, which is stable for our set of equations. The maximum
number of space discretization points that we have used is $10^{5}$ and the
time propagation is achieved by using a step $\delta t=0.4$. The initial
values for the densities are digitized, i.e. they take only the values $0.0$
and $1.0$ at the different space locations according to the chosen particle
density. As in the Monte Carlo simulations, we restrict ourselves to a
particle density of $0.5$, and choose $\gamma=0.1$. To define a cluster,
we set a threshold value $\phi _{th}=0.3$
for the density of the holes, i.e. the point $x$ where $\phi (x)$ crosses $%
\phi _{th}$ from above (below) marks the beginning (end) of a cluster. To
compute the average cluster size $\bar{l}_{c}(t)$, we average over $4000$
independent random initial configurations, using the same method as in the
Monte Carlo simulations.

Following the evolution of a typical initial condition, we can distinguish
several stages. Within the first few time steps, the sharp
distinction between initially filled ($\phi (x,0)=0$) and initially empty ($%
\phi (x,0)=1$) regions blurs. Domains of positive density ($\rho ^{+}(x,0)=1$%
) leak rapidly to higher $x$-values until they encounter domains of negative
density. At $t\sim 100$, larger domains of high mass density have formed,
typically covering $15-20$ discretization points.
Even though the mass density in these clusters is not always
fully saturated, almost all of these regions are included in the
cluster count, due to the fairly high threshold value, $\phi _{th}=0.3$.
Considering the two particle densities, we observe that the left 
(right) edges of clusters are marked by sharp peaks in 
$\rho ^{+}$ ($\rho ^{-}$).
Highly saturated domains are characterized by two peaks which
are almost perfect mirror images of one another. The traveler regions 
are still quite inhomogeneous and evolve rapidly. At later times 
(e.g., $t\sim 1000$), these traveler regions have become much more uniform, 
and essentially all clusters have a mass density
near unity.
Moreover, it appears that the density profiles
inside the clusters develop a $\tan $-shape reminiscent of the stationary
solution, Eq.~(\ref{rho+}). 
In other words, the late-time profiles can be described as being 
patched together from alternating traveller and cluster domains, each of
which seems nearly ``equilibrated'' within itself but not yet in
``equilibrium'' with the others. We will return to these observations below.

\begin{figure}[tbp]
\begin{center}
\epsfig{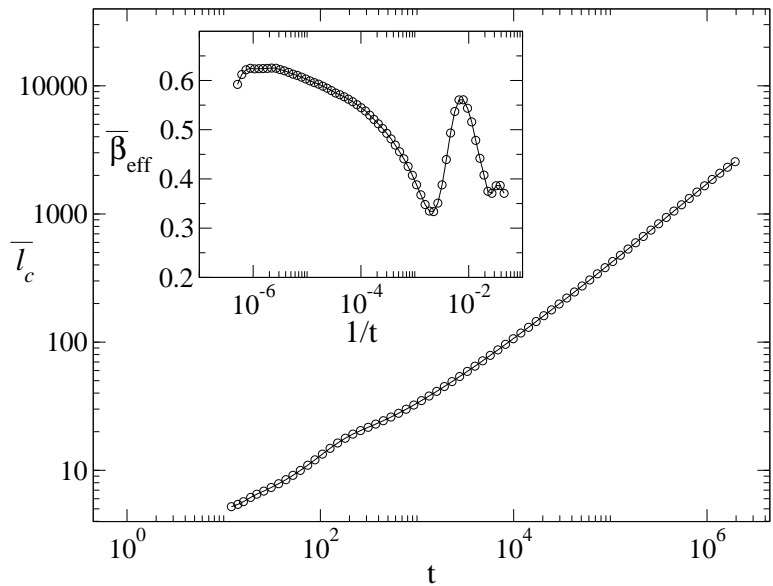}
\caption{The average cluster size $\bar{l}_{c}(t)$ and the effective 
growth exponent
$\beta_{eff}$ for $\gamma=0.1$ from mean-field numerical simulations.   }
\label{Growth_exp_ODE}
\end{center}
\vspace{-0.5cm}
\end{figure}

Our results for the average cluster size, $\bar{l}_{c}(t)$, and for the
local growth exponent $\bar{\beta}_{eff}\equiv d\ln (\bar{l}_{c}(t))/d(\ln t)
$ are shown in Fig.~\ref{Growth_exp_ODE}. At early times ($t\sim
100$), the local growth exponent develops a pronounced peak which can be
traced to the rapid formation of partially unsaturated clusters. Not
surprisingly, the detailed behavior of $\bar{\beta}_{eff}$ in this regime is
very sensitive to the chosen threshold $\phi _{th}$. A lower $\phi _{th}$
will reduce the size of this peak. For later times, the mean-field $\bar{%
\beta}_{eff}$ behaves very similarly to its Monte Carlo counterpart. It
slowly approaches a limiting value $0.63\pm 0.01$ before the
system reaches steady state. As in the MC simulations, the limiting value
``creeps up'' as the number of space discretization points (the system size)
increases, so that the mean-field results appear to be are consistent with a 
growth exponent of $2/3$.

Let us conclude this section by discussing a few simple analytic attempts
towards computing the growth exponent. None of them generate a satisfactory
estimate of the growth exponent, whence we conclude that its origin may be
quite subtle. Taking our clues from the Monte Carlo simulations, we note
that particles spend only a negligeable amount of time in transit between
clusters. Moreover, the density of travellers remains approximately constant
in time. It seems therefore reasonable to focus on the sizes (lengths) of the
clusters only. These evolve as neighboring clusters exchange particles.
Starting from the strictly one-dimensional coarse-grained configurations
described in Section (\ref{MC}), the simplest representation of a
sequence ($i=1,2,...$ ) of clusters is to specify their lengths, $l_{i}$, 
and the net positive and negative particle currents, $j^{\pm
}(l_{i})$, which flow through them. Assuming that cluster $i$ gains positive
(negative)\ particles from its left (right) neighbor and loses them to its
right (left)\ neighbor, we may write 
\begin{eqnarray}
\dot{l}_{i}=j^{+}(l_{i-1})+j^{-}(l_{i+1})-\left[ j^{+}(l_{i})+j^{-}(l_{i})%
\right] 
\nonumber
\end{eqnarray}%
Clearly, an overall constant cancels on the right hand side, so that only
the $l$-dependent \emph{excess} currents contribute. If we assume that the
cluster interiors are nearly ``equilibrated'' in the latest stages of the
growth process and invoke the mean-field result $j^{+}_{ex}(l_{i})\simeq
j^{-}_{ex}(l_{i})\propto 1/l_{i}^{2}$, simple scaling gives us $\dot{l}\propto
1/l^{2}$ or $l\sim t^{1/3}$. This clearly produces a poor estimate ($1/3$)
for the growth exponent. A somewhat better -- but still unsatisfactory --
estimate can be obtained if we replace the mean-field expression for the
currents by its exact counterpart. To do so, we recall that the cluster
interiors are almost entirely free of holes so that the dynamics inside the
clusters is dominated by $+/-$ exchanges, following a totally asymmetric
simple exclusion process (TASEP)\ \cite{Schuetz}. The associated excess
currents are exactly known:\ $j^{+}(l_{i})=j^{-}(l_{i})\propto 1/l_{i}$.
Simple scaling now yields $l\sim t^{1/2}$ which is faster than the
mean-field estimate but still slower than the observed $t^{2/3}$ behavior.
We note briefly that the same exponent ($\beta =1/2$) results from a
representation of the cluster growth process in terms of coalescing but
otherwise noninteracting random walkers \cite{Mettetal,Athens}. We believe
that the deficiencies of these approaches can be traced, by monitoring the
evolution of our clusters, to the rapid and systematic disintegration of
small ones. While this bias towards larger clusters can be modelled in an ad
hoc (but quantitatively very accurate) fashion by introducing appropriate
interactions between the walkers \cite{Mettetal}, we have as yet no reliable
insight into the physical origin of these interactions.

Finally, let us offer an intuitive picture which may provide some hints 
towards a growth exponent of $2/3$. 
First, the strongest evidence for this value
results from simulations with small $\gamma $. This suggests a time scale
separation between the dynamics of the traveller domains, dominated by fast
particle-hole exchanges, and the cluster regions, controlled by charge
exchanges on a much slower time scale set by $\gamma ^{-1}$. This is clearly
reflected in the mean-field currents, Eqs.~(\ref{MFT-trav}) and (\ref%
{MFT-cluster}). In other words, particles spend only very brief periods
travelling from one cluster to the next, compared to the time required to
traverse a large cluster from one end to the other. Moreover, the
microscopic dynamics within the interior of clusters is a simple TASEP which
can be described, under very well-defined conditions \cite{LPS}, by the
deterministic Burgers equation \cite{Burgers}. It is therefore tempting to
neglect the traveller regions and to think of the sequence of clusters in
terms of coarsening Burgers shocks. In this context, it is well known \cite%
{Burgers} that the average distance of shocks grows as $t^{2/3}$, provided
an average over random initial conditions is taken. Of course, this
reasoning remains a conjecture until a more rigorous analytic formulation
can be found.

\section{Conclusions}

To summarize, we have analyzed a coarsening phenomenon in a two-species
driven diffusive system on an $L\times 2$ periodic lattice. Positive
(negative)\ particles hop in the positive (negative)\ $x$-direction with
rate $1$, provided the destination site is empty. If it is filled by a
particle of opposite charge, the exchange occurs with rate $\gamma <1$.
Cross-lane exchanges are unbiased. Since $\gamma <1$, two opposing particles
impede each other, resulting in a temporary local blockage. For sufficiently
small $\gamma $, or sufficiently high overall particle density, these local
jams aggregate (``coarsen'') into a macroscopic cluster which characterizes
the final steady state of finite systems. A simple interpretation of this
dynamics is traffic of fast and slow cars on a two-lane ring road, viewed
from a co-moving frame.

Using Monte Carlo simulations and the numerical integration of a set of
mean-field equations, we study how an initially random distribution of
particles evolves to this inhomogeneous steady state, i.e., we quantify the
coarsening process in which local blockages form larger and larger clusters.
Measuring the average cluster size, $l_{c}(t)$, as a function of time, both
methods indicate power law growth $l_{c}(t)\sim t^{\beta }$ with a growth
exponent $\beta \simeq 2/3$ for the late stages of growth. In fact, Monte
Carlo simulations show excellent data collapse for small $\gamma $ and a
range of system sizes if $l_{c}(t)/L$ is plotted vs $t/L^{3/2}$. However, as 
$\gamma $ increases, the scaling exponents have to be adjusted to achieve a
comparable quality of data collapse. We interpret this as further evidence
for the observation that the inhomogeneous stationary state will not survive
in the thermodynamic limit \cite{Kafri,Kafri1,Georgiev}.

Since the stationary properties of the $L\times 1$ system can be rigorously
mapped \cite{Kafri,Kafri1} into an exactly solvable one-dimensional
zero-range process (ZRP)\ \cite{ZRP} involving one species of particles, it
is natural to compare our findings to recent results for the coarsening
dynamics of such models \cite{ZRP-dyn1}. They are defined on a chain of
sites, each of which can be occupied by an integer number, $k=0,1,2,...$ of
particles. The hopping rate $w(k)$ of a given particle depends only on the
number, $k$, of particles located at the originating site. Under
well-defined conditions on the $w\left( k\right) $, the density in the bulk
is limited to a critical value, and excess particles accumulate
(``condense'')\ at a non-extensive fraction of sites \cite{ZRP}. Cluster
sites are then defined as those sites where a macroscopic fraction of
particles has accumulated \cite{ZRP-dyn1}. As the system approaches the
final stationary state, one monitors how the average number of particles at
these cluster sites grows with time. Two basic universality classes can be
distinguished:\ If the $w(k)$ are symmetric, i.e., particles hop to the
right and left with equal rates, the dynamic exponent is found to be $z=3$,
resulting in a growth exponent $\beta =1/3$. In contrast, if the particle
hopping is biased, the growth occurs faster, with $z=2$ and $\beta =1/2$.
Clearly, both of these are incompatible with our results. We note that
growth exponents of $\beta =2/3$ can be ``engineered'' in one-dimensional
ZRPs involving \emph{two} species of particles \cite{ZRP-dyn2}, but only for
very special rates which break a fundamental symmetry of our dynamics,
mainly, CP invariance. Hence, we believe that there is little to learn from
these results for our case. Clearly, the fast coarsening observed here,
compared to growth exponents of $1/3$ or $1/2$ indicates that some essential
characteristics of our dynamics cannot be captured in the simple ZRP
picture. At this point, we can only speculate what these features might be.
\newline \\
\textit{Acknowledgments.} We thank G. Korniss, J. Krug, M. Evans, 
Y. Kafri, D. Mukamel, and G. Sch\"{u}tz for helpful 
discussions.
This work is partially supported by the NSF through DMR 0414122.

\end{document}